\documentclass[aps,prl,twocolumn,superscriptaddress]{revtex4-1}

\usepackage[utf8]{inputenc}
\usepackage[T1]{fontenc}
\usepackage[british]{babel}
\usepackage{xcolor,amsmath,amssymb,amsthm,amsfonts,dsfont,physics}
\definecolor{darkred}{rgb}{0.5,0,0}
\definecolor{darkgreen}{rgb}{0,0.5,0}
\definecolor{darkblue}{rgb}{0,0,0.5}
\usepackage[colorlinks,breaklinks,linkcolor=darkred,citecolor=darkgreen,urlcolor=darkblue]{hyperref}

\newcommand{\id}{\mathds{1}}

\newcommand{\md}{\mathrm{d}}
\newcommand{\rob}{\mathcal{R}}
\renewcommand{\leq}{\leqslant}
\renewcommand{\geq}{\geqslant}

\begin{document}

\author{S\'ebastien Designolle}
\thanks{These authors contributed equally to this work.}
\affiliation{Department of Applied Physics, University of Geneva, 1211 Geneva, Switzerland}
\author{Roope Uola}
\thanks{These authors contributed equally to this work.}
\affiliation{Department of Applied Physics, University of Geneva, 1211 Geneva, Switzerland}
\author{Kimmo Luoma}
\affiliation{Institut f\"ur Theoretische Physik, Technische Universit\"at Dresden, D-01062 Dresden, Germany}
\author{Nicolas Brunner}
\affiliation{Department of Applied Physics, University of Geneva, 1211 Geneva, Switzerland}

\date{\today}

\title{Set coherence: basis-independent quantification of quantum coherence}

\begin{abstract}
  The coherence of an individual quantum state can be meaningfully discussed only when referring to a preferred basis.
  This arbitrariness can however be lifted when considering sets of quantum states.
  Here we introduce the concept of set coherence for characterising the coherence of a set of quantum systems in a basis-independent way.
  We construct measures for quantifying set coherence of sets of quantum states as well as quantum measurements.
  These measures feature an operational meaning in terms of discrimination games and capture precisely the advantage offered by a given set over incoherent ones.
  Along the way, we also connect the notion of set coherence to various resource-theoretic approaches recently developed for quantum systems.
\end{abstract}

\maketitle

\textit{Introduction.---}
The superposition principle is a direct consequence of the linearity of quantum mechanics.
Given a set of orthogonal quantum states, their coherent superposition also represents a possible state.
The coherence stems from the fact that the phase relation between the various orthogonal states in the superposition is well defined.
This key concept of quantum theory has broad implications.
It is a central element for the existence of genuine randomness in quantum measurements and is also the basis of the phenomenon of entanglement.
Consequently, these ideas play a fundamental role in quantum information processing, quantum metrology, quantum transport, and many more important research directions.

A natural question is therefore to characterise the coherence of quantum systems, in particular for quantum states.
An intense research effort has been devoted to these questions in recent years, leading notably to the development of a general resource theory of quantum coherence; see, e.g., Refs~\cite{Abe06,BCP14,LM14,CG16,WY16,VS16,YMG+16,SAP17}.
There the coherence of a quantum state can be quantified via specific measures.
Interestingly, these measures have been shown to have an operational meaning, capturing precisely the advantage offered by a given quantum state (with coherence) for a certain task, compared to any possible incoherent quantum state~\cite{NBC+16,PCB+16}.
The case of quantum measurements has been investigated as well~\cite{BKB19,CGS+19}.

However, as intuition suggests, the above ideas can be meaningfully formalised only with respect to a preferred basis (or preferred reference frame).
Indeed, a single quantum state has intrinsically no coherence; for instance all pure quantum states are equivalent to each other if no basis is specified.
That is, the notion of quantum coherence for a single state is necessarily a relative property; it can be defined only with regard to a given reference.
While the choice of a preferred basis can be motivated in certain cases (for instance, choosing the energy basis in a thermodynamic setting), this basis dependence arguably limits the general scope and applicability of these ideas.

In this work, we follow a different approach for defining and quantifying the coherence of quantum systems in an absolute way, i.e., without referring to any preferred basis.
This provides a basis-independent (or reference-frame-independent) quantification of coherence.
The main idea is to consider a \textit{set of quantum states}, instead of a single state.
Consider for instance a pair of nonorthogonal pure states.
Clearly there exists no unitary that can map this pair of states to an orthogonal pair.
Hence, for any possible basis choice, the pair will necessarily feature some level of coherence.
More generally, given a set of states, a meaningful quantity can be defined by minimising the coherence of the states in the set over all possible basis choices. We term this quantity set coherence.
We note that a related approach was proposed in Ref.~\cite{HSS07}, where the authors developed an entropic measure for the quantumness of an ensemble of states, reflecting the entropy production in the ensemble.
In contrast our approach is motivated by the more recent resource-theoretic perspective, and thus tailored to the problem of characterising the intrinsic coherence of a set of states.

Below, we start by introducing formally the concept of set coherence. Then, we present two measures for set coherence of quantum states.
We derive explicit expressions for the set coherence of sets of qubit states and discuss sets featuring the highest set coherence.
This reveals a direct connection to the notion of commutativity.
Next, we investigate the notion of set coherence for quantum measurements.
Then, we demonstrate the operational meaning of our measures via quantum games.
In particular, the measure of set coherence of a set of states quantifies the advantage offered by this set over any incoherent set.
Finally, we conclude with a discussion on possible future directions.

\textit{Set coherence.---}
Consider a set of $n$ quantum states
\begin{equation}
  \vec{\varrho}=\{\varrho_j\}_{j=1}^n,
\end{equation}
where all states $\varrho_j$ are defined on a Hilbert space of dimension $d$.
While each state in the set has per se no absolute coherence (one can always express $\varrho_j$ in the basis in which it is diagonal), the situation can be different when considering the entire set of states $\vec{\varrho} $.
This motivates us to introduce a notion of set coherence characterising the coherence of the set $\vec{\varrho}$ in an absolute manner, i.e., without referring to any specific basis or reference frame.

To proceed, we follow a resource-theoretic approach.
Namely, we first define \emph{free} sets of states, that is, those featuring zero set coherence.
Intuitively, the latter consist of sets of states $\vec{\varrho}$, such that there exists a choice of basis (a unitary $U$) for which all states in the set $U \varrho_j U^\dagger$ become diagonal.
Formally, the free set is given by
\begin{equation}\label{eqn:free2}
  \mathcal{F}_n =\bigg\{\vec{\varrho}\,\,\,\bigg|\,\,\exists U,\,\forall j,\,U\varrho_jU^{\dagger}=\sum_{i=1}^d p(i|j)|i\rangle\langle i|\bigg\},
\end{equation}
where $p(\cdot|j)$ is a probability distribution for $j=1\ldots n$ and $\{\ket{i}\}_{i=1}^d$ denotes the computational basis.
If no such unitary can be found, then the set of states features nonzero set coherence.
Our goal is now to construct a measure for this effect, which is not straightforward due to the nonconvexity of the free set $\mathcal{F}_n$ (see Appendix~\hyperref[app:max]{A}).

We first consider the free set defined in Eq.~\eqref{eqn:free2} but restricting for the moment to a fixed unitary, taken for simplicity to be $U=\id$.
This corresponds to having a fixed reference basis, namely, the computational one; in due course, reference to this arbitrary choice of basis will disappear.
The free set is given by \cite{SAP17}
\begin{equation}\label{eqn:free}
  F_n=\bigg\{\vec{\varrho}\,\,\,\bigg|\,\,\forall j,\,\varrho_j=\sum_{i=1}^d p(i|j)|i\rangle\langle i|\bigg\}
\end{equation}
and is now convex, so that one can define the so-called generalised robustness of a given set of states $\vec{\varrho}$ with respect to $F_n$, i.e.,
\begin{equation}\label{eqn:Robustness}
  \rob_{F_n}(\vec{\varrho}) = \min\qty{t\geq 0\,\,\bigg|\,\, \frac{\vec{\varrho} + t\,\vec{\tau}}{1+t}\in F_n},
\end{equation}
where the optimisation is performed over all sets $\vec{\tau}$ with the same number of states and dimension as $\vec{\varrho}$.

To remove the dependency on a reference basis, we now minimise the above measure with respect to any possible basis choice.
Formally, we define the max-robustness of set coherence of $\vec{\varrho}$ as
\begin{equation}\label{eqn:setcohsum}
  \rob(\vec{\varrho})=\min_U\rob_{F_n}(U\vec{\varrho}\,U^\dagger).
\end{equation}
where the minimisation is performed over all unitaries $U$ acting on $\mathds{C}^d$.
Clearly, the above quantity is basis-independent and corresponds to the intrinsic (or minimal) amount of coherence present in the set.

While this measure captures a general property of the set $\vec{\varrho}$, one can nevertheless express it in terms of the robustness of the individual states $\varrho_j$ in the set.
More precisely, we show in Appendix~\hyperref[app:max]{A} that
\begin{equation}\label{eqn:sumrob}
  \rob(\vec{\varrho}) = \min_U  \max_j \rob_{F_1}(U\varrho_j\,U^\dagger),
\end{equation}
hence the name max-robustness.
This naturally suggests another possible measure for set coherence, replacing the maximum by the average over the states.
We thus define the mean-robustness of set coherence as
\begin{equation}\label{eqn:setcohsum1}
  \rob_1(\vec{\varrho})=\min_U
  \frac1n \sum_{j=1}^n \rob_{F_1}(U\varrho_j U^{\dagger}) .
\end{equation}

Importantly both $\rob$ and $\rob_1$ are faithful measures, in the sense that $\rob(\vec{\varrho})=0$ if and only if the set $\vec{\varrho}$ is incoherent, i.e., belongs to $\mathcal{F}_n$.
While these measures are not convex (see Appendix~\hyperref[app:max]{A}), similarly to the free set $\mathcal{F}_n$, we will see below that they nevertheless have a clear operational meaning.

In the following we will investigate these measures in various scenarios.
The max-robustness $\rob$ will be useful for discussing the set coherence of quantum measurements.
For sets of states, we will focus our attention mostly on the second measure $\rob_1$.
It turns out that $\rob_1$ is more convenient to calculate, as we will see below, and it provides a lower bound on $\rob$, as clearly $\rob_1(\vec{\varrho}) \leq \rob(\vec{\varrho})$.

\textit{Sets of qubit states.---}
We now illustrate the above ideas considering sets of qubit states ($d=2$).
We take advantage of the Bloch sphere representation: $\vec{q}_j$ denotes the Bloch vector of the state $\varrho_j$, i.e., $\varrho_j=(\id+\vec{q}_j\cdot\vec{\sigma})/2$, where $\vec{\sigma}$ contains the Pauli matrices $(\sigma_x,\sigma_y,\sigma_z)$.

To compute the set coherence measure $\rob_1(\vec{\varrho})$ for qubit sets, we proceed as follows.
Note first that the basis-dependent free set $F_n$, as defined in Eq.~\eqref{eqn:free}, now corresponds to sets of states all aligned with the vertical axis of the sphere (i.e., diagonal in the $\sigma_z$ basis).
As shown in Ref.~\cite{PCB+16} the robustness of each individual state reduces to the norm of its off-diagonal elements, i.e.,
\begin{equation}\label{eqn:robqubit}
  \rob_{F_1}(\varrho_j)=2\left|\langle0|\frac{\id+\vec{q}_j\cdot\vec{\sigma}}{2}|1\rangle\right|=\|\vec{q}_j\|\,|\sin{(\vec{e}_z,\vec{q}_j)}|,
\end{equation}
where $(\vec{e}_z,\vec{q}_j)$ is the angle between $\vec{q}_j$ and the $z$-axis.
With this, Eq.~\eqref{eqn:setcohsum1} simplifies to
\begin{equation}\label{eqn:setcohqubit}
  \rob_1(\vec{\varrho})=\frac1n\min_{\vec{p}\in S^2}\sum_{j=1}^n\,\|\vec{q}_j\|\,|\sin{(\vec{p},\vec{q}_j)}|,
\end{equation}
where the minimisation is now performed over unit-length vectors $\vec{p}$ on the Bloch sphere $S^2$.
The optimal vector $\vec{p}$ indicates the basis choice where the coherence of the set $\vec{\varrho}$ is minimised.
Note also that by using the relation $\|[\rho,\eta]\|=\frac{1}{2}\norm{\vec r}\norm{\vec v}\abs{\sin(\vec r,\vec v)}$, where $\vec r$ and $\vec v$ are the Bloch vectors of the states $\rho$ and $\eta$ respectively, one can rewrite
\begin{equation}
  \rob_1(\vec{\varrho})=\frac2n\min_{\ket{\psi}}\sum_{j=1}^n \big\|[\ketbra{\psi},\varrho_j]\big\|,
\end{equation}
where the optimisation is over pure qubit states.
Hence there is here a direct connection between set coherence and commutativity in the qubit case.

We start our analysis with sets of $n=2$ qubit states, the case $n=1$ trivially giving $\rob(\varrho_1)=0$ by aligning $\vec{p}$ with $\vec{q}_1$.
For pairs of pure qubit states the minimum in Eq.~\eqref{eqn:setcohqubit} is reached when $\vec{p}$ is either $\vec{q}_1$ or $\vec{q}_2$; hence
\begin{equation}\label{eqn:qubits}
  \rob_1(\vec{\varrho})= \frac12\big|\sin(\vec{q}_1,\vec{q}_2)\big|=\sqrt{\tr(\varrho_1\varrho_2)[1-\tr(\varrho_1\varrho_2)]}.
\end{equation}
Note that for pairs of mixed states, the optimal $\vec{p}$ is aligned with the Bloch vector of the purest $\varrho_j$ (i.e., the longest Bloch vector).
From the above equation, we see that, to maximise the set coherence, one should choose a pair of pure qubit states with Bloch vectors that are orthogonal, i.e., $|(\vec{q}_1,\vec{q}_2)|= \pi/2$.
For $n=2$ we thus get $\rob_1 \leq \rob_1^* :=\max_{\vec\varrho}\rob_1(\vec\varrho)= 1/2$.

More generally, we can characterise the sets of $n$ qubit states featuring the largest set coherence, see Appendix~\hyperref[app:most]{B} for details.
For triplets, i.e., $n=3$, one can show that ${\rob_1 \leq \rob_1^* = 2/3}$, the upper bound being attained when the three Bloch vectors form an orthonormal basis of $\mathds{R}^3$.
For some values of $n$, we can go further by using known results on optimisation over the sphere~\cite{Tho04,CK07}.
For $n=4$ one has ${\rob_1\leq\rob_1^*=1/\sqrt{2}}$, reached by states whose Bloch vectors form a regular tetrahedron, while for $n=6$ one has ${\rob_1\leq\rob_1^*=\sqrt{5}/3}$, obtained from half of an icosahedron.
The case of $n=5$ does not have any general answer in the literature and is notoriously hard~\cite{Sch13}.
Note that when $n \rightarrow \infty$ the optimal distribution of pure states tends to be uniform over the Bloch sphere and it follows that ${\rob_1 \leq \rob_1^*=\pi/4}$.
All these results are summarised in Table~\ref{tab:maxrob}.

We also discuss our alternative measure, namely, the max-robustness of set coherence.
As mentioned above, we have that $\rob_1(\vec{\varrho}) \leq \rob(\vec{\varrho})$, the inequality being tight only for incoherent sets of states.
For $n=2$, similarly to Eq.~\eqref{eqn:qubits} one can show that $\rob(\varrho_1,\varrho_2) = \sqrt{1-\tr(\varrho_1\varrho_2)}$ so that the maximal value is $\rob^*:=\max_{\vec\varrho}\rob(\vec\varrho)=1/\sqrt2$, also obtained for an orthogonal pair.
For $n=3$, it seems numerically that ${\rob^*=\sqrt{3}/2}$, obtained when the three vectors form a trine on an equator of the Bloch sphere.

\textit{Qudits.---}
Going beyond qubits using our methods turns out to be challenging, as the Bloch representation becomes more complex.
We can nonetheless still make a few statements.

First, observe that the case of a pair of pure states of arbitrary dimension corresponds to the qubit case, as the states span only a qubit subspace.
Therefore the rightmost expression in Eq.~\eqref{eqn:qubits} is applicable to any pair of pure states.
For $n>2$ and $d>2$ the situation is more complicated.
One can nevertheless prove from the inequality $\rob_{F_{1}}(\varrho)\leq d-1$ of Ref.~\cite{PCB+16} that
\begin{equation}\label{eqn:maxrobup}
  \rob^*\leq d-1\quad\mathrm{and}\quad\rob_1^*\leq\frac{(n-1)(d-1)}n.
\end{equation}
Although for $d=2$ the bound on $\rob$ is tight for $n\rightarrow\infty$ and the one on $\rob_1$ for $n=2,3$, we observe numerically that this is not the case in general.
Moreover, it seems that constructions based on mutually unbiased bases do not lead to the largest values of $\rob_1$.
For the case of sets of pure states, an interesting open question is whether the set coherence relates to properties of the Gram matrix (a matrix with entries given by the inner products of each pair of states in the set), which is known to identify uniquely its set of states, up to a unitary.

Finally, we note that lower bounds on the set coherence could be obtained by adapting the method developed in Ref.~\cite{CYJ+20}, where optimisation problems over unitaries can be relaxed to a hierarchy of semidefinite programs~\cite{BV04}.

\begin{table}[t]
  \centering
  \begin{tabular}{|c|c|c|}
    \hline
    $~~n~~$  & $ \rob_1^*$                         & Optimal sets                          \\ \hline
    2        & $\frac12$                           & Orthonormal pair (e.g., $X,Z$)        \\
    3        & $\frac23$                           & ~~Orthonormal basis (e.g., $X,Y,Z$)~~ \\
    4        & $\frac{1}{\sqrt2}\approx0.7071$     & Regular tetrahedron                   \\
    6        & $~~\frac{\sqrt5}{3}\approx0.7454~~$ & Half of an icosahedron                \\
    $\infty$ & $\frac{\pi}{4}\approx0.7854$        & Uniform distribution                  \\ \hline
  \end{tabular}
  \caption{
    Maximal values $\rob_1^*$ of the set coherence measure $\rob_1$ for sets of $n$ qubit states ($d=2$).
    The sets achieving these optimal values are also described by means of the geometry of their Bloch representation (see the main text for details).
  }
  \label{tab:maxrob}
\end{table}

\textit{Quantum measurements.---}
The notion of set coherence naturally applies to the case of quantum measurements (or positive operator-valued measures, POVMs).
Indeed the latter are represented by a set of Hermitian operators: $\mathcal A=\{A_a\}_{a=1}^n$ with the properties $A_a\geq0$ for all $a$ and $\sum_a A_a=\id$.
Note that the operators must sum up to the identity but do not need to have unit trace.

To quantify the set coherence of a POVM $\mathcal{A}$ we use the robustness $\rob(\mathcal{A})$ defined as in Eq.~\eqref{eqn:setcohsum} by replacing the set of density matrices $\{\varrho_j\}_{j=1}^n$ by an $n$-outcome POVM $\{A_a\}_{a=1}^n$.
The free set and basis-dependent robustness are constructed as in Eqs~\eqref{eqn:free} and \eqref{eqn:Robustness} with the only difference that all sets of operators considered, such as $\vec\tau$, should form a POVM.
The detailed construction of this measure, as well as its related mean-robustness, can be found in Appendix~\hyperref[app:robmeas]{C}.
In what follows, we concentrate on the measure constructed in line with Eq.~\eqref{eqn:setcohsum}.

We first discuss the set coherence for the case of a single POVM.
Clearly, we have that $\rob(\mathcal A)=0$ for all projective measurements, i.e., measurements with ${A_a^2=A_a}$ for all $a$, as the POVM elements commute with on another and can hence be simultaneously diagonalised.
Note that this is also the case for binary measurements.
There are however nonprojective POVMs that feature nonzero set coherence.
For qubit POVMs, we find numerically (similarly to the case of states; see Appendix~\hyperref[app:most]{B}) that the ternary ($n=3$) POVM with the most set coherence is the trine ($\rob= 1/\sqrt{3}$), while for $n=4$ we get $\rob = 1/\sqrt{2}$ for the symmetric informationally complete POVM (with Bloch vectors forming a regular tetrahedron).

It is relevant to comment on the relation between the set coherence of a POVM and some recently developed quantifiers of the usefulness of a POVM.
First, in Ref.~\cite{SL19} the authors quantify the informativeness of a POVM through its robustness with respect to those POVMs that have elements proportional to the identity operator.
The latter correspond to generalised coin-flip measurements, i.e., measurements with state-independent outcome distributions.
Clearly these POVMs have zero set coherence, and hence they belong to our free set.
This shows that any feasible point of the robustness measure for informativeness is also a feasible point in our optimisation.
This further implies that the robustness of informativeness upper bounds the set coherence.
There exist however POVMs that are informative but have no set coherence (e.g., projective measurements).
Second, one can consider POVMs that are simulable with projective measurements \cite{OGWA17,HQV+17}, i.e., measurements in the convex hull of projective measurements.
Any POVM with no set coherence is in this convex hull, as we show in Appendix~\hyperref[app:birkhoff]{D} based on a generalisation of the Birkhoff--von Neumann theorem~\cite{CLM+96}.
Hence, the robustness of a POVM with respect to the set of projective simulable measurements lower bounds the set coherence.
Finally, note that both the informativeness and nonprojective simulability have an operational interpretation in terms of a performance in a state discrimination task, as their corresponding free sets are convex~\cite{OGWA17,SL19,OB19,UKS+19}.
For set coherence, the free set is not convex although it is somewhere in between the two mentioned convex free sets.
In the next section we show that an interpretation through discrimination tasks is nevertheless possible.

We now move to the case of a set of POVMs.
In this case, one could expect a connection between the set coherence and the incompatibility of sets of POVMs as captured, e.g., via non-joint-measurability.
Joint measurability asks whether for a set of POVMs there exists a common POVM that functions as their common readout.
Clearly, zero set coherence guarantees mutual commutativity and hence joint measurability by using the product POVM $G_{\vec a}=\prod_x A_{a_x|x}$.
This is indeed a POVM because $[A_{a_x|x},A_{a_y|y}]=0$ and it has the property $A_{a|x}=\sum_{\vec{a}}\delta_{a_x,a}G_{\vec a}$; i.e., neglecting all but the outcome $a_x$ gives the POVM $\{A_{a_x|x}\}_{a_x}$.
Hence, the set coherence is an upper bound on the incompatibility robustness~\cite{UBGP15}.
However, there exist compatible sets of POVMs featuring nonzero set coherence, such as noisy $X$ and $Z$ measurements, and one may expect that a high enough set coherence ensures incompatibility.
These are interesting questions for future work.

\textit{Set coherence as a quantum game.---}
We now discuss the operational meaning of set coherence in terms of a quantum discrimination game.
We consider the case of a set of states $\vec{\varrho}$ and show that if $\rob_1(\vec{\varrho})>0$, then there exists a specific discrimination game for which $\vec{\varrho}$ provides an advantage over any incoherent set (i.e., sets of states that have zero set coherence).
Moreover, the value of $\rob_1(\vec{\varrho})$ quantifies precisely the relative advantage provided by $\vec{\varrho}$ over any incoherent set.

To construct such a game we consider a task of subchannel discrimination, i.e., distinguishing between different branches of a time evolution.
The branches are modelled as sets of completely positive maps $\mathcal C=\{\mathcal I_a\}_a$ with the property that $\sum_a\mathcal I_a$ is trace-preserving.
Given $\mathcal C$ and a final measurement $\mathcal A$, the goal is to identify which subchannel $\mathcal{I}_a$ has been applied.
The resource is the initial state $\varrho$.
The success probability is given by
\begin{align}
  p_{succ}(\varrho,\mathcal C,\mathcal A):=\sum_{a} \tr(\mathcal I_a(\varrho) A_a).
\end{align}
In the case in which one considers a fixed reference basis, the robustness measure $\rob_{F_1}$ quantifies the relative advantage offered by $\varrho$ over any state $\tau$ in the free set~\cite{NBC+16,PCB+16}, i.e.,
\begin{align}
  \label{eqn:Upperboundsingle}
  \frac{p_{succ}(\varrho,\mathcal C,\mathcal A)}{\max\limits_{\tau\in U^\dagger F_1U} p_{succ}(\tau,\mathcal C,\mathcal A)}
  \leq 1+\rob_{U^\dagger F_1U}(\varrho) .
\end{align}
This relation can be derived from Eq.~\eqref{eqn:Robustness} (with $n=1$) and holds for any unitary $U$ (specifying the reference frame), any set of subchannels $\mathcal C$, and any POVM $\mathcal A$.

Moving now to the case of a set of states $\vec{\varrho}$ and making the construction basis-independent by minimising over unitaries, one can show that
\begin{align}\label{eqn:RobAdv}
  \frac1n\min_U\sum_j\sup_{\mathcal C_j,\mathcal A_j}\frac{p_{succ}(\varrho_j,\mathcal C_j,\mathcal A_j)}{\max\limits_{\tau\in U^\dagger F_1U} p_{succ}(\tau,\mathcal C_j,\mathcal A_j)}= 1+ \rob_1(\vec\varrho).
\end{align}
All details of the derivation can be found in Appendix~\hyperref[app:games]{E}.
Note that here the discrimination procedure $(\mathcal C_j,\mathcal A_j)$ depends on $U$.
In other words, for any reference basis, i.e., for any given $U$, there exists a set of subchannel discrimination tasks $(\vec{\mathcal C},\vec{\mathcal A})$, in which the set of states $\vec\varrho$ outperforms any incoherent set in this reference frame, with relative advantage given by $\rob_{U^\dagger F_1U}(\vec\varrho)$.
When minimising this advantage over reference frames, one gets the mean-robustness of set coherence $\rob_1(\vec\varrho)$.

A similar construction can be made for our other measure of set coherence $\rob$.
In particular, this can also be adapted to the case of quantum measurements, where the set coherence robustness of a POVM quantifies the relative advantage in a state discrimination task.
We have sketched the proofs for these scenarios in Appendix~\hyperref[app:games]{E}.

We note that, in the case of measurements, the task-based interpretation sheds light on a natural question in quantum measurement theory.
Namely, the notion of commutativity of POVMs, i.e., the requirement that $[A_{a|1},A_{b|2}]=0$ for all $a,b$, is a type of measurement compatibility that lacks an operational interpretation.
Commutativity implies all known types of compatibility such as joint measurability~\cite{HMZ16}, unavoidable measurement disturbance~\cite{HW10}, and coexistence~\cite{LP97}, all of which can be given a task-oriented interpretation~\cite{WPF09,QVB14,UMG14,CHT19,SSC19,OB19,ULGP18,UVB19,UKS+19,TU20}.
It is clear that our notion of set coherence of measurements does not exhaust commutativity of POVMs, as one can easily construct a POVM that does not commute with itself but commutes with a trivial POVM.
However, for binary measurements our notion coincides with commutativity and, hence, in this scenario we get an operational interpretation of commutativity in the spirit of Eq.~\eqref{eqn:RobAdv}.

\textit{Conclusion.---}
We developed a notion of set coherence for characterising the coherence of a set of quantum systems.
This provides an approach for quantifying quantum coherence in a basis-independent manner.
This is appealing from the physical standpoint but becomes formally more challenging due to the nonconvexity of the resource theory.
Nevertheless, we showed that meaningful measures can be constructed for set coherence.
Some of these ideas could be useful for building resource theories for other quantum resources, such as non-Markovianity, as the set of Markovian channels is also known to be nonconvex~\cite{WECC08}.
In parallel, it would also be interesting to see if the present resource theory of set coherence can be ``convexified,'' for instance, by taking as the free set the convex hull of $\mathcal{F}_n$.

Finally, it would be interesting to investigate the relevance of set coherence in settings where sets of quantum states (or measurements) naturally appear, for instance in quantum key distribution (QKD) or quantum computation.
Could one design a secure QKD protocol based on any set of states featuring nonzero set coherence?
One may also consider quantifying the coherence of a quantum dynamical evolution, where a continuous set of states is explored over time.

\textit{Acknowledgements.---}
We thank Ryan Matzke and Dmitriy Bilyk for useful discussions and Micha\l{} Horodecki for pointing out the relevant work of Ref.~\cite{HSS07}.
RU is grateful for the financial support provided by the Finnish Cultural Foundation.
We acknowledge financial support from the Swiss National Science Foundation (Starting grant DIAQ and NCCR SwissMAP).

\bibliography{DULB21}
\bibliographystyle{sd2}

\appendix

\section{Appendix A: Properties of the max-robustness of set coherence}
\label{app:max}

In this Appendix we discuss some properties of the max-robustness of set coherence $\rob$, as well as the nonconvexity of the problem.

Let us start with the relation in Eq.~\eqref{eqn:sumrob}.
We first show that the joint robustness defined in Eq.~\eqref{eqn:Robustness} boils down to the maximum of the individual ones~\cite{PCB+16}, that is,
\begin{equation}\label{eqn:max}
  \rob_{F_n}(\vec{\varrho}) = \max_{j=1\ldots n} \rob_{F_1}(\varrho_j).
\end{equation}
This observation makes computation of this quantity easier and is proven below.
First, any feasible point of Eq.~\eqref{eqn:Robustness} consists of feasible points of the individual robustnesses.
Hence, Eq.~\eqref{eqn:Robustness} gives immediately an upper bound on the maximum of these robustnesses.
Going the other way, any set of feasible points of individual robustnesses can be converted into a feasible point of Eq.~\eqref{eqn:Robustness}, which we prove  for the case of pairs of states, the generalisation to arbitrary sets being straightforward.
Take some feasible points $t_1$ and $t_2$ and the corresponding $\tau_1$ and $\tau_2$ of the individual robustnesses.
Without loss of generality, we can assume that $t_1\geq t_2$.
Define $\tilde\tau_2=[(1+t_1)\varrho-\varrho_2]/t_1$, where $\varrho=(\varrho_2+t_2\tau_2)/(1+t_2)$.
Clearly $\tilde\tau_2$ is a quantum state as $t_1\geq t_2$.
Now the triplet $(\tau_1,\tilde\tau_2,t_1)$ is a feasible point of Eq.~\eqref{eqn:Robustness} as $\varrho$ has no coherence by definition.
This shows that the maximum of the individual robustnesses is an upper bound on Eq.~\eqref{eqn:Robustness}.
The desired relation Eq.~\eqref{eqn:sumrob} then follows.

Another important aspect is the nonconvexity of the free set $\mathcal{F}$.
As an example, one can consider the following two sets of states
\begin{align}
  &\vec{\varrho}_1 = \left\{ |0\rangle\langle0|,\frac{1}{3}|0\rangle\langle0|+\frac{2}{3}|1\rangle\langle1| \right\} \\
  &\vec{\varrho}_2 = \left\{ |+\rangle\langle+|,\frac{1}{4}|+\rangle\langle+|+\frac{3}{4}|-\rangle\langle-| \right\}
\end{align}
where $|0\rangle,|1\rangle,|+\rangle$ and $|-\rangle$ are the eigenstates of the Pauli matrices $\sigma_z$ and $\sigma_x$.
Each set clearly belongs to the free set $\mathcal{F}_n$.
However, their convex mixtures is not.
Note that these pairs of states can be transformed into pairs of binary POVMs by seeing a state $\varrho$ as one POVM element, the other one being $\id-\varrho$.
Hence, also the set of commuting pairs of POVMs is not convex.

\section{Appendix B: Sets of qubit states featuring the highest set coherence}
\label{app:most}

In this Appendix we prove that some sets of qubit states that have the highest set coherence among those with the same number of states can be found via a connection with known mathematical problems.
Specifically, for four, six, and twelve states, we show that qubit states whose Bloch vectors form a (regular) tetrahedron, an octahedron, and an icosahedron are optimal.

We start by recalling the formal problem that we want to solve and by giving elementary solutions for pairs and triplets of qubit states.
The idea is to find sets of qubit states with Bloch vectors $\vec{q}_1\ldots\vec{q}_n$ maximising the set coherence defined in Eq.~\eqref{eqn:setcohsum}, namely, thanks to Eq.~\eqref{eqn:setcohqubit},
\begin{equation}\label{eqn:maxminsumapp}
  \rob_1^*=\max_{\vec{q}_1\ldots\vec{q}_n}\,\,\,\min_{\vec{p}}\,\,\,\frac1n\sum_{i=1}^n\|\vec{q}_j\|\,|\sin(\vec{p},\vec{q}_j)|.
\end{equation}
First notice that the optimal states are obviously pure, so that we can take $\|\vec{q}_j\|=1$ in the following.
Then for pairs of states, we have already seen in the main text that the inner minimisation is reached when $\vec{p}$ is either of the (pure) states and that the maximum value $\rob_1^*=1/2$ is obtained for $n=2$ orthogonal states.
For triplets of states, the proof is quite similar: when taking $\vec{p}=\vec{q}_1$ we get the upper bound on Eq.~\eqref{eqn:maxminsumapp}
\begin{equation}
  \rob_1^*\leq\max_{\vec{q}_1\ldots\vec{q}_n}\,\,\,\frac{|\sin(\vec{q}_1,\vec{q}_2)|+|\sin(\vec{q}_1,\vec{q}_3)|}{3}\leq\frac23,
\end{equation}
which is trivially saturated for Bloch vectors forming an orthonormal basis so that $\rob_1^*=2/3$ for $n=3$ qubit states.

More generally, we can get $n$ upper bounds on Eq.~\eqref{eqn:maxminsumapp} by fixing $\vec{p}=\vec{q}_i$ for $i=1\ldots n$.
Then the mean of these upper bounds is still greater than $\rob_1^*$ so that
\begin{equation}\label{eqn:upsin}
  \rob_1^*\leq\max_{\vec{q}_1\ldots\vec{q}_n}\,\,\,\frac1{n^2}\sum_{i,j}|\sin(\vec{q}_i,\vec{q}_j)|.
\end{equation}
Note that the right-hand-side of Eq.~\eqref{eqn:upsin} now looks like an interaction energy where, up to a sign, the effective interaction between two points $\vec{q}_i$ and $\vec{q}_j$ is repulsive when the angle $(\vec{q}_i,\vec{q}_j)$ is smaller than $\pi/2$ but attractive for larger angles.
Then, computing the right-hand-side of Eq.~\eqref{eqn:upsin} amounts to placing $n$ points on the sphere $S^2$ so as to minimise the energy of the configuration.
This kind of problem has been considered initially by Thomson who was looking for the optimal placement of electrons (interacting repulsively through Coulomb's law) on their hypothetical spherical orbit around the nucleus~\cite{Tho04}.
Perhaps surprisingly exact solutions are only known for two, three, four, five, six, and twelve electrons, the proof for five being quite recent (2010) and relying heavily on computer assistance~\cite{Sch13}.
Generalisation to other energy functions and higher dimensions have also been investigated~\cite{CK07,BDM18}.

Going back to our problem, we note that the sinus function is actually symmetric around this angle $\pi/2$ so that we can restrict ourselves to the upper half of the sphere by reversing points on the lower half.
This does not affect the energy of the configuration and we are left with an interaction that is only repulsive.
Formally this amounts to working in the real projective space, for which results have been obtained in Ref.~\cite[Sec.~8]{CK07} for completely decreasing functions, that is, satisfying $(-1)^kf^{(k)}(x)\geq0$ for all nonnegative integer $k$, acting on the squared distance between points.
In our case, we can write
\begin{equation}
  |\sin(\vec{q}_i,\vec{q}_j)|=\|\vec{q}_i-\vec{q}_j\|\sqrt{1-\frac{\|\vec{q}_i-\vec{q}_j\|^2}{4}}
\end{equation}
so that, up to an affine transformation, the resulting function we want to minimise is indeed completely decreasing.
Then the solutions for $n=3$, $n=4$, and $n=6$ follow and are respectively given by half of an octahedron, that is, an orthonormal basis, a (regular) tetrahedron, and half of an icosahedron.
The corresponding values for $\rob_1^*$ are $2/3$ for $n=3$, already obtained above, $1/\sqrt{2}\approx0.7071$ for $n=4$, and $\sqrt{5}/3\approx0.7454$ for $n=6$.
Note that the full octahedron has a robustness of 4, which is optimal among \emph{symmetric} distributions of six points, that is, forming three pairs of diametrically aligned points, but not among \emph{all} distributions of six points, since the best value given above is reached for half of an icosahedron.
Finally, having a completely decreasing energy function allows directly to get that the asymptotic optimal distribution is indeed uniform, so that
\begin{equation}
  \rob_1 \leq \rob_1^* =\frac{1}{4\pi}\int_0^{2\pi}\!\!\!\int_0^\pi\sin^2\theta\md\theta\md\varphi=\frac{\pi}{4}\approx0.7854,
\end{equation}
as mentioned in the main text.

For the alternative measure $\rob$, the situation is more involved as the connection breaks down.
More precisely, because of the nonlinearity of the expression~\eqref{eqn:max}, the averaging procedure leading to Eq.~\eqref{eqn:upsin} does not work so that we cannot get an effective repulsive energy on the sphere.
Thus for this measure we were only able to get analytically $\rob^*=1/\sqrt2\approx0.7071$ in the case of $n=2$, for which the optimal basis of two pure states lies precisely on the bisector of the two Bloch vectors.
For higher $n$ we used a numerical method to estimate $\rob^*$ based on semi-definite programming~\cite{Lof04,mosek} and heuristic optimisation over unitaries \cite{SHH10}.

For $n=3$ the optimal set seems to be formed by pure qubit states with Bloch vectors forming an equilateral triangle on a great circle, reaching a value of $\rob^*=\sqrt{3}/2\approx0.8660$ with and optimal basis aligned with one of the states.
Note that triplets of states with Bloch vectors forming an orthonormal basis only feature a set coherence of $\sqrt{2/3}\approx0.8165$, obtained for a basis aligned with the diagonal of the cube defined by the three Bloch vectors.
For $n=4$ the optimal set is surprisingly neither the tetrahedron reaching $\sqrt{2/3}$ nor the square reaching $1/\sqrt{2}$.
The best value we got numerically is $\rob^*\gtrapprox0.91$.
For $n\rightarrow\infty$ it is clear that the uniform distribution achieves $\rob=1$, which saturates the bound of Eq.~\eqref{eqn:maxrobup} in the main text.
Therefore this distribution is also optimal for $\rob$.

\section{Appendix C: Set coherence for measurements}
\label{app:robmeas}

In the case of measurements, the construction of the free set and the robustness measures requires a few technical notes.
First, as we operate in the space of measurements or tuples thereof, the considered noise and free set consist of measurement assemblages.
The basis-dependent free set for $k$ measurements is defined as
\begin{equation}\label{eqn:freemeasreferencedep}
  \tilde F_k =\bigg\{\mathcal M\,\,\,\bigg|\,\, M_{a|x}=\sum_{i=1}^d \alpha(i|a,x)|i\rangle\langle i|\ \forall a,x\bigg\},
\end{equation}
where $\mathcal M=\{M_{a|x}\}$, $\alpha(i|a,x)\geq0$ and ${\sum_a\alpha(i|a,x)=1}$ for all $i$ and $x$.
This motivates the definition of the robustnesses $\rob$ of a measurement assemblage $\mathcal M$ with $n$ measurements as in the case of states, i.e.,
\begin{equation}\label{eqn:Robustnessmeasure}
  \rob_{\tilde F_k}(\mathcal M) = \min\qty{t\geq 0\,\,\bigg|\,\, \frac{\mathcal M + t\,\mathcal N}{1+t}\in \tilde F_k},
\end{equation}
where the optimisation is over all measurement assemblages $\mathcal N$ having the same dimension as well as the same number of inputs and outputs as $\mathcal M$.
Using this measure, we define
\begin{equation}\label{eqn:setcohsummeas}
  \rob(\mathcal M)=\min_U\rob_{\tilde F_k}(U\mathcal M\,U^\dagger).
\end{equation}
where the minimisation is performed over all unitaries $U$ acting on $\mathds{C}^d$.

In the case of measurements, one has to be careful when defining the mean-robustness $\rob_1$.
Specifically, the needed robustnesses of single measurements correspond to the noise tolerance of a whole measurement (when mixed with another whole measurement) with respect to the set $\tilde F_1$, i.e., the mixing is not performed on the level of single POVM elements.
Taking this into account, we define the mean-robustness for an assemblage $\mathcal M$ of $n$ measurements as
\begin{equation}\label{eqn:setcohsummeas2}
  \rob_1(\mathcal M)=\min_U
  \frac1k \sum_{x=1}^k \rob_{\tilde F_1}(U\{M_{a|x}\} U^{\dagger}).
\end{equation}
We note that in the case of single measurements, the robustnesses $\rob$ and $\rob_1$ coincide and can have nonzero values.
This behaviour differs from the case of single states.

\section{Appendix D: Incoherent measurements are projective simulable}
\label{app:birkhoff}

In this Appendix, we give the details of the proof that sets of measurements with no coherence can be simulated by means of projective measurements~\cite{OGWA17}.
In spite of its apparent simplicity, this problem is nontrivial and its proof relies on an extension of the Birkhoff--von Neumann theorem to rectangular matrices.

The first step is to notice that it suffices to solve it for single measurements, since the generalisation to sets of measurements follows naturally.
Then we take a single measurement $\mathcal{A}=\{A_a\}_a$ with $n$ (nontrivial) outputs in dimension $d$ and we consider it to be incoherent in the computational basis, namely,
\begin{equation}
  A_a=\sum_{i=1}^dp(i|a)\ketbra{i}.
\end{equation}
We want to show that there exist (i) positive coefficients $q_k$ summing to one and (ii) projective measurements $\{\{P_{a|k}\}_a\}_k$ such that
\begin{equation}
  A_a=\sum_kq_kP_{a|k}.
\end{equation}

For pedagogical reasons, we first assume that $n=d$ and $\tr A_a=1$ for all $a$.
Then the $d\times d$ matrix $M$ with $m_{ij}=p(i|j)$ is said to be doubly stochastic as it consists of positive entries summing up to one on all rows and all columns.
In virtue of the Birkhoff--von Neumann theorem, we can then deduce that $M$ is a convex combination of permutation matrices.
Since such matrices correspond to projective measurements onto the computational basis, the desired property follows.
In general, the proof is quite similar but relies on the extension of the above-mentioned theorem to rectangular matrices.

Specifically, now we do not make any specific assumption on $n$ or $\tr A_a$.
The matrix $M$ is then of size $d\times n$ and is such that the sum over all rows is one whereas the sum over the $j$th column is $\tr A_j$.
Denoting $t=\min_a\tr A_a\ne0$, we make all sums over columns equal by substracting $(\tr A_j-t)/d$ to all entries $m_{ij}$.
In terms of measurements, this amounts to starting the decomposition by means of trivial (projective) measurements being zero for all outcomes except one, namely,
\begin{equation}
  A_a=\sum_{j=1}^n\frac{\tr A_j-t}{d}\delta_{a,j}\id+\frac{t}{d}\sum_{i=1}^d\tilde{p}(i|a)\ketbra{i},
\end{equation}
where the choice of $t$ ensures that $\tilde{p}(i|a)\geq0$ for all $i$ and $a$.
Then the matrix $\tilde{M}$ with nonnegative coefficients $\tilde{m}_{ij}=\tilde{p}(i|j)$ satisfies the hypothesis of Ref.~\cite{CLM+96}, that is, the sum over the rows is $n$ and over the columns $d$.
We can decompose $\tilde{M}$ into extremal matrices of this (convex) set (with the same sums over rows and columns) and these are known to have integer entries, see, e.g., Ref.~\cite[Proposition~1(iii)]{CLM+96}.
Specifically, this means that
\begin{equation}
  \tilde{p}(i|a)=\sum_k\lambda_kr_k(i|a),
\end{equation}
where $\lambda_k$ are positive coefficients summing up to one and $r_k(i|a)$ are nonnegative integers.
Note that the number of terms in this decomposition can be bounded thanks to Carathéodory's theorem.

The matrices associated to $r_k(i|a)$ can further be decomposed into $n$ matrices corresponding to projective measurements, simply by repeatedly taking off one to any positive integer of each row in order to create a matrix with exactly one coefficient one per row, which is precisely one of a projective measurement.
The equality of the sums over all rows guarantees that this procedure will indeed be possible.
For example, we get
\begin{equation}
  \begin{pmatrix}3&&\\1&2&\\&2&1\\&&3\end{pmatrix}=
  \begin{pmatrix}1&&\\1&&\\&1&\\&&1\end{pmatrix}+
  \begin{pmatrix}1&&\\&1&\\&1&\\&&1\end{pmatrix}+
  \begin{pmatrix}1&&\\&1&\\&&1\\&&1\end{pmatrix},
\end{equation}
where the above $4\times3$ matrices correspond to diagonal measurements with $n=3$ outcomes in dimension $d=4$, the columns being the diagonals of the different outcomes.
More formally, we define, for $0\leq l\leq n-1$,
\begin{equation}
  s_k^{(l)}(i|a):=\delta_{a,\min\left\{b\,\big|\,r_k^{(l)}(i|b)\neq0\right\}},
\end{equation}
where $r_k^{(0)}:=r_k$ and
\begin{equation}
  r_k^{(l+1)}(i|a):=r_k^{(l)}(i|a)-s_k^{(l)}(i|a),
\end{equation}
so that $\sum_ar_k^{(l)}(i|a)=n-l$ and $\sum_as_k^{(l)}(i|a)=1$ for all $l$.
With this last equality and the fact that $s_k^{(l)}(i|a)^2=s_k^{(l)}(i|a)$ one can see that the measurements $S_{a|k}^{(l)}:=\sum_is_k^{(l)}(i|a)\ketbra{i}$ are indeed projective.

All in all, we have brought the matrix $M$ corresponding to $\mathcal{A}$ to a ``doubly stochastic'' form $\tilde{M}$ and then expressed the extremal points of the set of such matrices in terms of projective measurements.
This means that $\mathcal{A}$ can indeed be decomposed as a convex combination of projective measurements, namely,
\begin{equation}
  A_a=\sum_{j=1}^n\frac{\tr A_j-t}{d}T_{a|j}+\frac{t}{d}\sum_k\lambda_k\sum_{l=0}^{n-1}S_{a|k}^{(l)},
\end{equation}
where $T_{a|j}=\delta_{a,j}\id$ are trivial projective measurements.

\section{Appendix E: Interpretation of the robustness measures as discrimination tasks}
\label{app:games}

We divide this Appendix into three sections.
We first present the task-oriented interpretations of the two different robustness measures for the case of sets of states.
Finally, we discuss the case of POVMs.

\subsection{Sets of states with the mean-robustness \texorpdfstring{$\rob_1$}{}}

Here we present the details of the construction outlined in the main text.
We consider a task of subchannel discrimination.
The subchannels are denoted $\mathcal C=\{\mathcal I_a\}$ with the property that $\sum_a\mathcal I_a$ is trace-preserving.
For a given resource state $\varrho$, the success probability is given by
\begin{align}
  p_{succ}(\varrho, \mathcal C,\mathcal A):=\sum_{a} \tr(\mathcal I_a(\varrho) A_{a}).
\end{align}
and the measure connects the relative advantage via
\begin{align}
  \label{eqn:Upperboundsingle2}
  \frac{p_{succ}(\varrho,\mathcal C,\mathcal A)}{\max\limits_{\tau\in U^\dagger F_1U} p_{succ}(\tau,\mathcal C,\mathcal A)}
  \leq 1+\rob_{U^\dagger F_1U}(\varrho) .
\end{align}

Moving now to the case of a set of states $\vec{\varrho}$ and minimising over unitaries we get
\begin{align}
  \label{eqn:Upperbound}
  \frac1n\min_U\sum_j\frac{p_{succ}(\varrho_j,\mathcal C_j,\mathcal A_j)}{\max\limits_{\tau_j\in U^\dagger F_1U} p_{succ}(\tau_j,\mathcal C_j,\mathcal A_j)}\leq 1+ \rob_1(\vec\varrho).
\end{align}
The above inequality can be turned into an equality by following methods developed in Ref.~\cite{NBC+16}.
The robustness measure can be cast as a conic program, the dual of which reads
\begin{eqnarray}\label{eqn:RobDual}
  1+\rob_{U^\dagger F_1U}(\varrho) = & ~\max~      & \tr(\varrho Y^U)                        \\
                                     & \text{s.t.} & Y^U\geq0                       \nonumber\\
                                     &             & \tr(Y^U T)\leq 1               \nonumber\\
                                     &             & \forall\ T\in U^\dagger F_1 U. \nonumber
\end{eqnarray}
This objective function is closely related to a specific subchannel discrimination task, following, e.g., the derivation of Ref.~\cite{NBC+16}.
Define a set of unitary operators $\{U_{a}\}_a$ that shifts the elements $\{e_k\}$ of the eigenbasis of $Y^U$ by $a$, i.e., $U_{a}e_k=e_{k+a}$.
Now define a set of subchannels $\mathcal I_{a}(\varrho)=\frac{1}{d}U_a\varrho U_a^\dagger$ and the discriminating POVMs $A_a = U_a Y^U U_a^\dagger/\tr(Y^U)$.
Hence for a given $U$ any term in the sum in Eq.~\eqref{eqn:RobDual} is, up to a scaling by $\tr(Y^U_j)$, an instance of a subchannel discrimination task.
As this scaling does not affect the ratios on the left-hand-side of Eq.~\eqref{eqn:Upperbound}, one has that
\begin{align}\label{eqn:RobAdv3}
  \frac1n\min_U\sum_j\sup_{\mathcal C_j,\mathcal A_j}\frac{p_{succ}(\varrho_j,\mathcal C_j,\mathcal A_j)}{\max\limits_{\tau_j\in U^\dagger F_1U} p_{succ}(\tau_j,\mathcal C_j,\mathcal A_j)}= 1+ \rob_1(\vec\varrho).
\end{align}
Note that here the discrimination procedure $(\mathcal C_j,\mathcal A_j)$ depends on $U$.
In other words, for any reference frame, i.e., for any given $U$, there exists a set of subchannel discrimination tasks $(\vec{\mathcal C},\vec{\mathcal A})$, in which the set of states $\vec\varrho$ outperforms those with no coherence in this reference frame with relative advantage given by $n+\sum_j\rob_{U^\dagger F_1U}(\varrho_j)$.
When minimising this advantage over reference frames and dividing by $n$, one gets the mean-robustness of set coherence $1+\rob_1(\vec\varrho)$.

\subsection{Sets of states with the max-robustness \texorpdfstring{$\rob$}{}}

As for the measure $\rob_1$, we need the dual formulation of $\rob_{F_n}$.
This reads
\begin{eqnarray}\label{eqn:RobDual2}
  1+\rob_{U^\dagger F_nU}(\vec\varrho) = & ~\max~      & \sum_j\tr(\varrho_j Y_j^U)              \\
                                         & \text{s.t.} & Y^U\geq0                       \nonumber\\
                                         &             & \tr(Y^U T)\leq 1               \nonumber\\
                                         &             & \forall\ T\in U^\dagger F_n U, \nonumber
\end{eqnarray}
where $Y^U=\oplus_j Y_j^U$.
To interpret this dual as a game, one can follow the exact same procedure as in the main text to define the subchannels and the discriminating POVMs for each $i$.
The only difference is, that in this case we need prior probabilities for the states, i.e., $p(i):=\tr(Y_j^U)/\tr(\sum_j Y_j^U)$, and we consider games with such probabilities fixed.

To define the game, we note that the above procedure gives a subgame $\mathcal G_j:=(p(j),\mathcal C_j,\mathcal A_j)$ for every $j$.
The full game is characterised by these subgames, i.e., $\mathcal G:=\{\mathcal G_j\}_j$.
The score $\mathcal S$ a set of states $\vec\varrho$ gives in the game $\mathcal G$ is defined as
\begin{align}
  \mathcal S(\vec\varrho,\mathcal G):=\sum_{a,j} p(j)\tr(\mathcal I_{a|j}(\varrho_j)A_{a|j}).
\end{align}
Solving $\vec\varrho$ from Eq.~\eqref{eqn:Robustness} shows in line with Eq.~\eqref{eqn:Upperbound} that
\begin{align}
  \label{eqn:Upperbound2}
  \frac{\mathcal S(\vec\varrho,\mathcal G)}{\max\limits_{\vec\tau\in U^\dagger F_nU} \mathcal S(\vec\tau,\mathcal G)}\leq 1+ \rob_{U^\dagger F_nU}(\vec\varrho).
\end{align}
As the scaling by $\tr(\sum_j Y_j^U)$ coming from the dual formulation does not affect the ratio on the left-hand-side of Eq.~\eqref{eqn:Upperbound2}, we conclude that
\begin{align}
  \label{eqn:Robadv2}
  \min_U\sup_{\mathcal G}\frac{\mathcal S(\vec\varrho,\mathcal G)}{\max\limits_{\vec\tau\in U^\dagger F_nU} \mathcal S(\vec\tau,\mathcal G)}= 1+\min_U\rob_{U^\dagger F_nU}(\vec\varrho).
\end{align}

\subsection{Sets of measurements with \texorpdfstring{$\rob$}{} (and \texorpdfstring{$\rob_1$}{})}

We recall from Appendix~\hyperref[app:robmeas]{C} that the robustness $\rob_{\tilde F_k}(\mathcal M)$ of a measurement assemblage $\mathcal M$ is defined as
\begin{equation}\label{eqn:Robustnessmeasure2}
  \rob_{\tilde F_k}(\mathcal M) = \min\qty{t\geq 0\,\,\bigg|\,\, \frac{\mathcal M + t\,\mathcal N}{1+t}\in \tilde F_k},
\end{equation}
where the optimisation is over all measurement assemblages $\mathcal N$ having the same dimension as well as the same number of inputs and outputs as $\mathcal M$.
The dual of this robustness with a given reference frame $U$ reads
\begin{eqnarray}\label{eqn:RobDualmeas}
  1+\rob_{U^\dagger \tilde F_kU}(\mathcal M) = & ~\max~      & \sum_{a,x}\tr(A_{a|x}Y_{a|x}^U)                \\
                                               & \text{s.t.} & Y^U\geq0                              \nonumber\\
                                               &             & \tr(Y^U T)\leq 1                      \nonumber\\
                                               &             & \forall\ T\in U^\dagger \tilde F_k U, \nonumber
\end{eqnarray}
where $Y^U=\oplus_{a,x}Y_{a|x}^U$.
Writing
\begin{equation}
  Y_{a|x}=\tr(Y_{a|x})\frac{Y_{a|x}}{\tr(Y_{a|x})}=:\tr(Y_{a|x})\varrho_{a|x}
\end{equation}
and
\begin{equation}
  p(a|x):=\frac{\tr(Y_{a|x})}{\sum_b\tr(Y_{b|x})}
\end{equation}
shows that, for each $x$, Eq.~\eqref{eqn:RobDualmeas} corresponds, up to a scaling factor of $\tr(\sum_b Y_{b|x})$, to a minimum error state discrimination probability for the set of states $\{p(a|x)\varrho_{a|x}\}_a$ with the measurements $\{A_{a|x}\}_a$.
Going one step further, one can define
\begin{equation}
  p(x)=\frac{\sum\limits_b \tr(Y_{b|x})}{\sum\limits_{b,y}\tr(Y_{b|y})}
\end{equation}
and note that Eq.~\eqref{eqn:RobDualmeas} corresponds (again up to a factor) to a minimum error state discrimination probability with prior information on the used set $x$.
The success probability in a minimum error discrimination task of a state assemblage $\mathcal{T}=\{\varrho_{a|x}\}_{a,x}$ with prior information and with a measurement assemblage $\mathcal M=\{A_{a|x}\}_{a,x}$ is defined as
\begin{align}
  p_{succ}(\mathcal{T},\mathcal M):=\sum_{a,x}p(x)p(a|x)\tr(\varrho_{a|x}A_{a|x}).
\end{align}
Solving $\mathcal M$ from Eq.~\eqref{eqn:Robustnessmeasure2} shows that
\begin{align}
  \label{eqn:Upperboundmeas}
  \frac{p_{succ}(\mathcal{T},\mathcal M)}{\max\limits_{\mathcal O\in U^\dagger \tilde F_kU}p_{succ}(\mathcal{T},\mathcal O)}\leq 1+ \rob_{U^\dagger \tilde F_kU}(\mathcal M).
\end{align}
It is then clear from the dual formulation that
\begin{align}
  \label{eqn:Robadvmeas}
  \min_U\sup_{\mathcal{T}}\frac{p_{succ}(\mathcal{T},\mathcal M)}{\max\limits_{\mathcal O\in U^\dagger \tilde F_kU}p_{succ}(\mathcal{T},\mathcal O)}= 1+ \rob(\mathcal M).
\end{align}

As the above calculation gives to the mean-robustness a task-interpretation, simply by repeatedly using the result on $\rob_{\tilde F_1}$, we do not explicitly spell out this case.

\end{document}